\documentclass [11pt,a4paper] {article}

\usepackage[cp1252]{inputenc}

\usepackage{amssymb}
\usepackage{amsmath}
\usepackage{amsfonts,amssymb}
\usepackage[dvips]{graphicx}
\usepackage{bbm}
\usepackage{enumerate}

\usepackage{amsthm}

\usepackage{cancel}

\DeclareMathAlphabet{\mathpzc}{OT1}{pzc}{m}{it}

\begin{document}

\title{Classical limit of quantum propositions}

\author{Arkady Bolotin\footnote{$Email: arkadyv@bgu.ac.il$\vspace{5pt}} \\ \textit{Ben-Gurion University of the Negev, Beersheba (Israel)}}

\maketitle

\begin{abstract}\noindent Contrary to classical semantics, the disjunction of two experimental propositions relating to pure states of a quantum system (“quantum propositions” for short) can be true even in the case where neither disjunct is true. This suggests that in such case either both disjuncts are false and so the distributive laws are not applicable to quantum propositions (this inference is accepted in quantum logic) or the disjuncts are not bivalent, i.e., neither true nor false, therefore the principle of bivalence is not applicable to quantum propositions. But, to accept the latter inference, one must explain how quantum propositions become bivalent in the classical limit. This paper shows the emergence of bivalence through the interaction between a quantum system and its environment and compares the environmentally induced bivalence with the classical limit of quantum logic.\\

\end{abstract}

\section{Introduction}  

\noindent The following observation suggests that classical logic is not applicable to experimental (i.e., verifiable or at least potentially falsifiable) propositions relating to pure states of quantum systems.\\

\noindent Consider the experimental propositions ``Spin along the $z$-axis is $+\frac{\hbar}{2}\,$'', ``Spin along the $x$-axis is $+\frac{\hbar}{2}\,$'' and ``Spin along the $x$-axis is $-\frac{\hbar}{2}\,$'' relating to pure states of a qubit – i.e., a two-state quantum-mechanical system (such as a one-half spin particle, say, an electron) – which can be denoted as $P_{z+}$, $P_{x+}$ and $P_{x-}$, respectively. Suppose that the qubit is prepared in the pure state described by the normalized eigenvector $|\Psi_{z+}\rangle = [\begin{smallmatrix} 1 \\ 0 \end{smallmatrix}]$ corresponding to the eigenvalue $+1$ of the Pauli matrix $\sigma_z$. This fact implies that in the state $|\Psi_{z+}\rangle$ the proposition $P_{z+}$ has the value of the truth. Provided a proposition is identified with the set of interpretations making the proposition true, the subspace $\mathcal{H}_{z+} = \{ a \in \mathbb{R},\,[\begin{smallmatrix} a \\ 0 \end{smallmatrix}]\}$ that includes the state $|\Psi_{z+}\rangle$ can be viewed as the representative of $P_{z+}$.\\

\noindent On the other hand, this state can be presented as $|\Psi_{z+}\rangle = \frac{1}{\sqrt{2}} (|\Psi_{x+}\rangle +|\Psi_{x-}\rangle)$, i.e., the superposition of the normalized eigenvectors  corresponding to the eigenvalues $+1$ and $-1$ of the Pauli matrix $\sigma_x$, namely, $|\Psi_{x+}\rangle = \frac{1}{\sqrt{2}} [\begin{smallmatrix} 1 \\ 1 \end{smallmatrix}]$ and $|\Psi_{x-}\rangle = \frac{1}{\sqrt{2}} [\begin{smallmatrix} \text{ }\text{ }1 \\ -1 \end{smallmatrix}]$. Each of these eigenvectors lies in the matching closed linear subspace, namely, $\mathcal{H}_{x+} = \{ a \in \mathbb{R},\,[\begin{smallmatrix} a \\ a \end{smallmatrix}]\}$ or $\mathcal{H}_{x-} = \{ a \in \mathbb{R},\,[\begin{smallmatrix} \text{ }\text{ }a \\ -a \end{smallmatrix}]\}$, meaning that in the states $|\Psi_{x+}\rangle \in \mathcal{H}_{x+}$ and $|\Psi_{x-}\rangle \in \mathcal{H}_{x-}$ the propositions $P_{x+}$ and $P_{x-}$ have the value of the truth in that order.\\

\noindent As $[\begin{smallmatrix} 1 \\ 0 \end{smallmatrix}] \notin \{ a \in \mathbb{R},\,[\begin{smallmatrix} a \\ a \end{smallmatrix}]\}$ as well as $[\begin{smallmatrix} 1 \\ 0 \end{smallmatrix}] \notin \{ a \in \mathbb{R},\,[\begin{smallmatrix} \text{ }\text{ }a \\ -a \end{smallmatrix}]\}$, one can observe that though the disjunction $P_{x+} \vee P_{x-}$ must be true, neither $P_{x+}$ nor $P_{x-}$ is true in the state $|\Psi_{z+}\rangle$.\\

\noindent A conclusion following from this observation is that both $P_{x+}$ and $P_{x-}$ are false in $|\Psi_{z+}\rangle$. But then, in this state the statement $P_{z+} \wedge (P_{x+}\vee P_{x-})$ must be true while the statement $(P_{z+} \wedge P_{x+}) \vee (P_{z+} \wedge P_{x-})$ must be false. Consequently, the distributive law of $\wedge$ over $\vee$ must be no longer valid in the logic of experimental quantum propositions. Such conclusion has been accepted in quantum logic of Birkhoff and von Neumann \cite{Birkhoff}.\\

\noindent However, another conclusion can be drawn from the above observation: it is that in the state $|\Psi_{z+}\rangle$ both $P_{x+}$ and $P_{x-}$ are undetermined, i.e., neither true nor false. And so, the principle of bivalence (stating that a proposition can only be described as either true or false \cite{Beziau}) must be no longer valid in the logic of experimental quantum propositions. Specifically, this logic must have \textit{supervaluation semantics} in which a disjunction assumes a definite truth value even when its disjuncts do not (i.e., its disjuncts have \textit{truth values gaps}) \cite{Varzi, Keefe}. But, to accept that the principle of bivalence is not applicable to experimental quantum propositions, it is necessary to explain how these propositions become bivalent in the classical limit (unless, of course, one has proof that our, i.e., classical, logic needs to be replaced by a non-standard logic). That is, it is essential to show how at some time or point in the course of the quantum-to-classical transition of the logical structure of experimental propositions a gappy (supervaluation) semantics turns into a gapless (bivalent) semantics.\\

\noindent The emergence of bivalent semantics in the logic of experimental quantum propositions is shown in the present paper.\\

\section{Algebraic structure of supervaluation logic}  

\noindent Just as a Heyting algebra is an algebraic structure of intuitionistic logic \cite{Ghilardi} and a Hilbert lattice is an algebraic structure of quantum logic (of Birkhoff and von Neumann) \cite{Ptak}, the collection of invariant-subspace lattices with non-intertwined nontrivial subspaces is an algebraic structure of supervaluation logic.\\

\noindent To show this, some preliminaries are in order first.\\

\noindent Let $\hat{P}$ be the projection operator on the Hilbert space $\mathcal{H}$. Then, \textit{the range} of this operator denoted by $\mathrm{ran}(\hat{P})$ is a closed linear subspace of $\mathcal{H}$; explicitly, $\mathrm{ran}(\hat{P})$ is the subset of the vectors $|\Psi\rangle \in \mathcal{H}$ that are in the image of $\hat{P}$, i.e.,\smallskip

\begin{equation} \label{RAN} 
   \mathrm{ran}(\hat{P})
   =
   \left\{
      |\Psi\rangle \in \mathcal{H}
      \text{:}
      \;
      \hat{P} |\Psi\rangle =  |\Psi\rangle
   \right\}
   \;\;\;\;  .
\end{equation}
\smallskip

\noindent Dually, $\mathrm{ker}(\hat{P})$ stands for the kernel of $\hat{P}$ (a closed linear subspace of $\mathcal{H}$ as well), i.e., the subset of the vectors $|\Psi\rangle \in \mathcal{H}$ that are mapped to zero by $\hat{P}$, namely,\smallskip

\begin{equation} \label{KER} 
   \mathrm{ker}(\hat{P})
   =
   \mathrm{ran}(\hat{1} - \hat{P})
   =
   \left\{
      |\Psi\rangle \in \mathcal{H}
      \text{:}
      \;
      (\hat{1} - \hat{P}) |\Psi\rangle =  |\Psi\rangle
   \right\}
   \;\;\;\;  ,
\end{equation}
\smallskip

\noindent where $\hat{1}$ stands for the identity operator. For that reason, the projection operator $\hat{1} - \hat{P}$ can be understood as \textit{the negation} of $\hat{P}$, i.e.,\smallskip

\begin{equation}  
   \neg\hat{P}
   =
   \hat{1} - \hat{P}
   \;\;\;\;  .
\end{equation}
\smallskip

\noindent As consequences of (\ref{RAN}) and (\ref{KER}), one has\smallskip

\begin{equation}  
   \mathrm{ran}(\hat{P})
   +
   \mathrm{ran}(\neg\hat{P})
   =
   \mathrm{ran}(\hat{1})
   =
   \mathcal{H}
   \;\;\;\;  ,
\end{equation}

\begin{equation}  
   \mathrm{ran}(\hat{P})
   \cap
   \mathrm{ran}(\neg\hat{P})
   =
   \mathrm{ran}(\hat{0})
   =
   \{0\}
   \;\;\;\;  ,
\end{equation}
\smallskip

\noindent where $\cap$ is the set-theoretical intersection, $\hat{0}$ is the zero operator, and the subsets $\{0\}$ and $\mathcal{H}$ are \textit{the trivial subspaces} of $\mathcal{H}$ (which correspond to the trivial projection operators $\hat{0}$ and $\hat{1}$, respectively).\\

\noindent A subspace ${\mathcal{H}}^{\prime} \subseteq \mathcal{H}$ is called \textit{an invariant subspace} under the projection operator $\hat{P}$ if\smallskip 

\begin{equation}  
   \hat{P}
   \text{:}
   \;
   {\mathcal{H}}^{\prime}
   \to
   {\mathcal{H}}^{\prime}
   \;\;\;\;  .
\end{equation}
\smallskip

\noindent This means that the image of every vector $|\Psi\rangle$ in ${\mathcal{H}}^{\prime}$ under $\hat{P}$ remains within ${\mathcal{H}}^{\prime}$ which can be denoted as\smallskip

\begin{equation}  
   \hat{P}\,{\mathcal{H}}^{\prime}
   =
   \left\{
      |\Psi\rangle \in {\mathcal{H}}^{\prime}
      \text{:}
      \;\;
      \hat{P}|\Psi\rangle
   \right\}
   \subseteq
   {\mathcal{H}}^{\prime}
   \;\;\;\;  .
\end{equation}
\smallskip

\noindent It is easy to see that $\hat{P}\,\mathrm{ran}(\hat{P}) \subseteq \mathrm{ran}(\hat{P})$ and $\hat{P}\,\mathrm{ran}(\neg\hat{P}) \subseteq \mathrm{ran}(\neg\hat{P})$ in addition to $\hat{P}\, \mathrm{ran}(\hat{1}) \subseteq \mathrm{ran}(\hat{1})$ and $\hat{P}\,\mathrm{ran}(\hat{0}) \subseteq \mathrm{ran}(\hat{0})$.\\

\noindent Let $\mathcal{L}(\hat{P})$ denote the set of the invariant subspaces in $\mathcal{H}$ i\textit{nvariant under the projection operator} $\hat{P}$:\smallskip

\begin{equation}  
   \mathcal{L}(\hat{P})
   =
   \left\{
      {\mathcal{H}}^{\prime} \subseteq \mathcal{H}
      \text{:}
      \;\;
      \hat{P}\,{\mathcal{H}}^{\prime}
      \subseteq
      {\mathcal{H}}^{\prime}
   \right\}
   \;\;\;\;  .
\end{equation}
\smallskip

\noindent Recall that the set of two or more nontrivial projection operators on $\mathcal{H}$ is called \textit{a context} $\Sigma$\smallskip

\begin{equation}  
   \Sigma
   =
   \left\{
      {\hat{P}}^{\,\prime}
      ,
      {\hat{P}}^{\,\prime\prime}
      ,
      \dots
   \right\}
   \;\;\;\;   
\end{equation}
\smallskip

\noindent if the next two conditions hold:\smallskip

\begin{equation}  
      {\hat{P}}^{\,\prime}
      ,
      {\hat{P}}^{\,\prime\prime}
      \in
      \Sigma
      \;\;
      \implies
      \;\;
      {\hat{P}}^{\,\prime}
      {\hat{P}}^{\,\prime\prime}
      =
      {\hat{P}}^{\,\prime\prime}
      {\hat{P}}^{\,\prime}
      =
      \hat{0}
   \;\;\;\;  ,
\end{equation}

\begin{equation}  
   \sum_{\hat{P} \in \Sigma}
   \hat{P}
   =
   \hat{1}
   \;\;\;\;  .
\end{equation}
\smallskip

\noindent Consider the set of the invariant subspaces $\mathcal{L}(\Sigma)$ \textit{invariant under each} $\hat{P} \in \Sigma$:\smallskip

\begin{equation}  
   \mathcal{L}(\Sigma)
   =
   \bigcap_{\hat{P} \in \Sigma}
   \mathcal{L}(\hat{P})
   \;\;\;\;  .
\end{equation}
\smallskip

\noindent The elements of this set form a complete lattice called \textit{the invariant-subspace lattice} of the context $\Sigma$ \cite{Radjavi}. By way of illustration, for the qubit, the invariant-subspace lattice of the context $\Sigma_{q}$, where $q \in \{z,x,y\}$, is\smallskip

\begin{equation}  
   \mathcal{L}(\Sigma_{q})
   =
   \mathcal{L}(\hat{P}_{q+})
   \cap
   \mathcal{L}(\hat{P}_{q-})
   =
   \left\{
      \mathrm{ran}(\hat{0})
      ,\,
      \mathrm{ran}(\hat{P}_{q+})
      ,\,
      \mathrm{ran}(\hat{P}_{q-})
      ,\,
      \mathrm{ran}(\hat{1})
   \right\}
   \;\;\;\;  .
\end{equation}
\smallskip

\noindent The lattice operations on $\mathcal{L}(\Sigma)$ are defined in an ordinary way: Particularly, \textit{the meet} $\wedge$ and \textit{the join} $\vee$ are defined by\smallskip

\begin{equation}  
   {\mathcal{H}}^{\prime}
   ,
   {\mathcal{H}}^{\prime\prime}
   \in
   \mathcal{L}(\Sigma)
   \;\;
   \implies
   \;\;
   \left\{
      \begin{array}{l}
         {\mathcal{H}}^{\prime}
         \wedge
         {\mathcal{H}}^{\prime\prime}
         =
         {\mathcal{H}}^{\prime}
         \cap
         {\mathcal{H}}^{\prime\prime}
         \in
         \mathcal{L}(\Sigma)
         \\ 
         {\mathcal{H}}^{\prime}
         \vee
         {\mathcal{H}}^{\prime\prime}
         =
         \left(
            ({\mathcal{H}}^{\prime})^{\perp}
            \cap
            ({\mathcal{H}}^{\prime\prime})^{\perp}
         \right)^{\perp}
         \in
         \mathcal{L}(\Sigma)
      \end{array}
   \right.
   \;\;\;\;  ,
\end{equation}
\smallskip

\noindent where $(\cdot)^{\perp}$ stands for the orthogonal complement of $(\cdot)$.\\

\noindent It is straightforward to see that each invariant-subspace lattice only contains the subspaces belonging to the mutually commutable projection operators, that is, each $\mathcal{L}(\Sigma)$ is a Boolean algebra. It follows then that for $\mathrm{ran}({\hat{P}}^{\,\prime}), \mathrm{ran}({\hat{P}}^{\,\prime\prime}) \in \mathcal{L}(\Sigma)$, one has\smallskip

\begin{equation} \label{ORT} 
      \mathrm{ran}({\hat{P}}^{\,\prime})
      \wedge
      \mathrm{ran}({\hat{P}}^{\,\prime\prime})
      =
      \{0\}
      \;\;
      \iff
      \;\;
      {\hat{P}}^{\,\prime}
      {\hat{P}}^{\,\prime\prime}
      =
      {\hat{P}}^{\,\prime\prime}
      {\hat{P}}^{\,\prime}
      =
      \hat{0}
   \;\;\;\;  ,
\end{equation}

\begin{equation}  
      \mathrm{ran}({\hat{P}}^{\,\prime})
      \wedge
      \mathrm{ran}({\hat{P}}^{\,\prime\prime})
      =
      \mathrm{ran}({\hat{P}}^{\,\prime})
      \;\;
      \iff
      \;\;
      {\hat{P}}^{\,\prime}
      {\hat{P}}^{\,\prime\prime}
      =
      {\hat{P}}^{\,\prime\prime}
      {\hat{P}}^{\,\prime}
      =
      {\hat{P}}^{\,\prime}
   \;\;\;\;  ,
\end{equation}

\begin{equation}  
      \mathrm{ran}({\hat{P}}^{\,\prime})
      +
      \mathrm{ran}({\hat{P}}^{\,\prime\prime})
      +
      \dots
      =
      \mathcal{H}
      \;\;
      \iff
      \;\;
      {\hat{P}}^{\,\prime}
      +
      {\hat{P}}^{\,\prime\prime}
      +
      \dots
      =
      \hat{1}
   \;\;\;\;  .
\end{equation}
\smallskip

\noindent Recall that two contexts are called \textit{intertwined} if they share one or more common elements \cite{Svozil}. As any intertwined context has at least one \textit{individual} element (a projection operator that is not shared by other contexts), each $\mathcal{L}(\Sigma)$ has a nonempty set of the individual subspaces (that are not shared by the lattices of other contexts).\\

\noindent Let $\mathcal{O} = \{\Sigma\}$ be the set of the contexts associated with the given quantum system. This set is in one-to-one correspondence with the collection of the lattices $\mathcal{L}(\Sigma)$ denoted $\{\mathcal{L}(\Sigma)\}_{\Sigma \in \mathcal{O}}$.\\

\noindent If $\mathrm{ran}(\hat{P}) \in \mathcal{L}(\Sigma)$ while $\mathrm{ran}(\hat{Q})$ is the individual subspace belonging to the different lattice from the collection $\{\mathcal{L}(\Sigma)\}_{\Sigma \in \mathcal{O}}$, then $\mathrm{ran}(\hat{P})$ and $\mathrm{ran}(\hat{Q})$ cannot meet each other within the structure of $\{\mathcal{L}(\Sigma)\}_{\Sigma \in \mathcal{O}}$. In symbols,\smallskip

\begin{equation}  
   \mathrm{ran}(\hat{P})
   \in
   \mathcal{L}(\Sigma)
   ,\,
   \mathrm{ran}(\hat{Q})
   \notin
   \mathcal{L}(\Sigma)
   \;
   \implies
   \;
   \mathrm{ran}(\hat{P})
   \;\cancel{\;\wedge\;}\;
   \mathrm{ran}(\hat{Q})
   \;\;\;\;  ,
\end{equation}
\smallskip

\noindent where the cancelation of the meet $\wedge$ indicates that this operation cannot be defined on such subspaces (recall that the meet is defined as an operation on pairs of elements from one lattice \cite{Davey}).\\

\noindent To bring experimental propositions relating to the pure states of the quantum system into contact with the lattice structure imposed on the closed linear subspaces of system's Hilbert space $\mathcal{H}$, let us construe such propositions in the following way.\\

\noindent Suppose that the system is prepared in the pure state $|\Psi\rangle$ residing in the closed linear subspace $\mathcal{H}_{|\Psi\rangle} \subseteq \mathcal{H}$, and let $\mathcal{H}_{P}$ be the nontrivial closed linear subspace of $\mathcal{H}$ that represents the experimental proposition $P$. Then, in the state $|\Psi\rangle$ the proposition $P$ should assume the value of truth or falsity in accordance with the propositional function $\mathrm{Prop}$, explicitly,\smallskip

\begin{equation} \label{PROP} 
   P(|\Psi\rangle)
   =
   \mathrm{Prop}
   \left(
      |\Psi\rangle
      \in
      \mathcal{H}_{|\Psi\rangle}
      \wedge
      \mathcal{H}_{P}
   \right)
   \;\;\;\;  .
\end{equation}
\smallskip

\noindent Take, for example, the case in which the subspaces $\mathcal{H}_{|\Psi\rangle}$ and $\mathcal{H}_{P}$ are members of one invariant-subspace lattice $\mathcal{L}(\Sigma)$. If, in addition to that, $\mathcal{H}_{|\Psi\rangle} \wedge \mathcal{H}_P = \mathcal{H}_{|\Psi\rangle}$, then, according to the above formula, the proposition $P$ must be \textit{a tautology}:\smallskip

\begin{equation}  
   P(|\Psi\rangle)
   =
   \mathrm{Prop}
   \Big(
      |\Psi\rangle
      \in
      \mathcal{H}_{|\Psi\rangle}
   \Big)
   \;\;\;\;  .
\end{equation}
\smallskip

\noindent On the other hand, if $\mathcal{H}_{|\Psi\rangle} \wedge \mathcal{H}_P = \mathcal{H}^{\prime}$, where $\mathcal{H}^{\prime}$ is the nontrivial closed linear subspace not equal to $\mathcal{H}_{|\Psi\rangle}$, then the condition $|\Psi\rangle \in \mathcal{H}_{|\Psi\rangle}$ entails either $|\Psi\rangle \in \mathcal{H}^{\prime}$ or $|\Psi\rangle \notin \mathcal{H}^{\prime}$. Consequently, in the state $|\Psi\rangle \in \mathcal{H}_{|\Psi\rangle}$ the proposition $P$ is expected to be either true or false. In a semantics defined by \textit{the bivaluation relation} (i.e., the function $b$ from the set of propositions into the set $\{0,1\}$ of bivalent truth values), this can be expressed equivalently (using 0 for \textit{false} and 1 for \textit{true}) as:\smallskip

\begin{equation}  
   b
   \left(
      P(|\Psi\rangle)
   \right)
   =
   b
   \bigg(
      \mathrm{Prop}
      \Big(
         |\Psi\rangle
         \in
         \mathcal{H}^{\prime}
      \Big)
   \bigg)
   \in
   \{0,1\}
   \;\;\;\;  .
\end{equation}
\smallskip

\noindent Then again, if $\mathcal{H}_{|\Psi\rangle} \wedge \mathcal{H}_P = \{0\}$, the proposition $P$ must be \textit{a contradiction}\smallskip

\begin{equation}  
   b
   \left(
      P(|\Psi\rangle)
   \right)
   =
   b
   \bigg(
      \mathrm{Prop}
      \Big(
         |\Psi\rangle
         \in
         \{0\}
      \Big)
   \bigg)
   =
   0
   \;\;\;\;   
\end{equation}
\smallskip

\noindent since any state of the quantum system $|\Psi\rangle$ (meaningful from the physical point of view) differs from 0.\\

\noindent It follows from this that in the state $|\Psi\rangle \in \mathrm{ran}(\hat{P}^{\,\prime})$, where $\hat{P}^{\,\prime} \in \Sigma$, only one proposition represented by the range of the projection operator belonging to the context $\Sigma$ can be assigned the value 1:\smallskip

\begin{equation}  
   \sum_{{\hat{P}}^{\,\prime\prime} \in  \Sigma}
   b
   \left(
      \mathrm{Prop}
      \left(
         |\Psi\rangle
         \in
         \mathrm{ran}({\hat{P}}^{\,\prime})
         \wedge
         \mathrm{ran}({\hat{P}}^{\,\prime\prime})
      \right)
   \right)
   =
   1
   \;\;\;\;  .
\end{equation}
\smallskip

\noindent By contrast, consider the case where the proposition $Q$ is represented by the nontrivial subspace $\mathcal{H}_{Q}$ that is not a member of the lattice $\mathcal{L}(\Sigma)$ containing the subspace $\mathcal{H}_{|\Psi\rangle}$. Because within the structure of $\{\mathcal{L}(\Sigma)\}_{\Sigma \in \mathcal{O}}$ the meet operation on $\mathcal{H}_{|\Psi\rangle} \in \mathcal{L}(\Sigma)$ and $\mathcal{H}_Q \notin \mathcal{L}(\Sigma)$ is not defined, the propositional function $\mathrm{Prop}$ turns out to be undetermined. Hence, according to the formula (\ref{PROP}), the proposition $Q$ cannot assume any value in the state $|\Psi\rangle \in \mathcal{H}_{|\Psi\rangle}$, which can be expressed in a bivalent semantics as:\smallskip

\begin{equation}  
   b
   \Big(
      Q(|\Psi\rangle)
   \Big)
   =
   b
   \bigg(
      \mathrm{Prop}
      \Big(
         |\Psi\rangle
         \in
         \mathcal{H}_{|\Psi\rangle}
         \;\cancel{\;\wedge\;}\;
         \mathcal{H}_{Q}
      \Big)
   \bigg)
   =
   \frac{0}{0}
   \;\;\;\;  ,
\end{equation}
\smallskip

\noindent where $\frac{0}{0}$ denotes an indeterminate value.\\

\noindent In this way, the nonexistence of the meet operation on pairs of the nontrivial ranges that do not lie in a common lattice $\mathcal{L}(\Sigma) \in \{\mathcal{L}(\Sigma)\}_{\Sigma \in \mathcal{O}}$ corresponds to truth-value gaps.\\

\noindent Assume that the trivial subspace $\mathcal{H} = \mathrm{ran}(\hat{1}) = \mathrm{ran}(\hat{Q}) + \mathrm{ran}(\neg\hat{Q}) $ represents the disjunction of the proposition $Q$ and its negation $\neg Q$:\smallskip

\begin{equation} \label{ASM} 
   Q
   \vee
   \neg Q
   \;
   \iff
   \;
   \mathcal{H}
   \;\;\;\;  .
\end{equation}
\smallskip

\noindent Then, in accordance with (\ref{PROP}) one finds:\smallskip

\begin{equation}  
   Q \vee \neg Q
   \:
   \big(|\Psi\rangle\big)
   =
   \mathrm{Prop}
   \big(
      |\Psi\rangle
      \in
      \mathcal{H}_{|\Psi\rangle}
      \wedge
      \mathcal{H}
   \big)
   \;\;\;\;  .
\end{equation}
\smallskip

\noindent As $\mathcal{H}_{|\Psi\rangle} \wedge \mathcal{H} = \mathcal{H}_{|\Psi\rangle}$, the propositional function $\mathrm{Prop}$ is evaluated to the truth in any allowable state $|\Psi\rangle$; therefore, the disjunction $Q \vee \neg Q$ is always true\smallskip

\begin{equation}  
   b
   \Big(
      Q \vee \neg Q
      \:
      \big(|\Psi\rangle\big)
   \Big)
   =
   1
   \;\;\;\;  ,
\end{equation}
\smallskip

\noindent even when both $Q$ and $\neg Q$ (represented by the subspaces $\mathrm{ran}(\hat{Q})$ and $\mathrm{ran}(\neg\hat{Q})$) are undetermined, i.e., $b\big(Q(|\Psi\rangle)\big) = \frac{0}{0}$ and $b\big(\neg Q(|\Psi\rangle)\big) = \frac{0}{0}$.\\

\noindent Hence, equipped with the assumption (\ref{ASM}), the gappy logic whose algebraic structure is given by the collection $\{\mathcal{L}(\Sigma)\}_{\Sigma \in \mathcal{O}}$ can be identified as \textit{supervaluationism}.\\

\section{Environmentally induced bivalence}  

\noindent Consider the following simple model. Let the qubit $S$ interact with its environment $E$ described by a collection of $N$ other qubits. Suppose that each environmental qubit has the preferred set of states (say, due to the design of the experiment), namely, $\{|\Psi_{kz±}\rangle\}$ where $ k \in \{1, \dots, N\}$, while the qubit $S$ has no preferred states. Correspondingly, the Hilbert space of the $k^{\text{th}}$ environmental qubit $\mathcal{H}_k$ is\smallskip

\begin{equation}  
   \mathcal{H}_k
   =
   \mathrm{ran}(\hat{1})
   =
   \sum_{\beta = \pm}
   \mathrm{ran}(\hat{P}_{kz\beta})
   \;\;\;\;  ,
\end{equation}
\smallskip

\noindent at the same time as the Hilbert space $\mathcal{H}_S$ of the qubit $S$ can be presented as\smallskip

\begin{equation}  
   \mathcal{H}_S
   =
   \mathrm{ran}(\hat{1})
   =
   \sum_{\alpha = \pm}
   \mathrm{ran}(\hat{P}_{Su^{\,\prime}\alpha})
   =
   \dots
   =
   \sum_{\alpha = \pm}
   \mathrm{ran}(\hat{P}_{Su^{\,\prime\prime}\alpha})
   =
   \dots
   \;\;\;\;  ,
\end{equation}
\smallskip

\noindent where $u^{\,\prime}$, $u^{\,\prime\prime}$, $\dots$ are arbitrary axes. Thus, the Hilbert space of the composite system $SE$ is the tensor product\smallskip

\begin{equation}  
   \mathcal{H}_{SE}
   =
   \mathrm{ran}(\hat{1})
   =
   \mathcal{H}_{S}
   \bigotimes_{k=1}^{N}
   \mathcal{H}_{k}
   \;\;\;\;  ,
\end{equation}
\smallskip

\noindent or, explicitly,\smallskip

\begin{equation}  
   \mathcal{H}_{SE}
   =
   \sum_{\beta = \pm}
   \dots
   \left(
      \sum_{\beta = \pm}
      \left(
         \sum_{\beta = \pm}
         \mathcal{H}_{S}
         \otimes
         \mathrm{ran}({\hat{P}}_{1z\beta})
      \right)
      \otimes
      \mathrm{ran}({\hat{P}}_{2z\beta})
   \right)
   \dots
   \otimes
   \mathrm{ran}({\hat{P}}_{Nz\beta})
   \;\;\;\;  .
\end{equation}
\smallskip

\noindent Let the structure of the invariant-subspace lattices be imposed on the closed linear subspaces of $\mathcal{H}_S$ and $\mathcal{H}_k$. One can observe then that even though the subspaces $\mathrm{ran}(\hat{P}_{Su^{\,\prime}\alpha})$ and $\mathrm{ran}(\hat{P}_{Su^{\,\prime\prime}\alpha})$ are not elements of the same invariant-subspace lattice, they will reside together in one of the invariant-subspace lattices containing the subspaces of the Hilbert space $\mathcal{H}_{SE}$.\\

\noindent To see this, consider spin of the qubit $S$ along the axes $z$ and $x$. The ranges $\mathrm{ran}(\hat{P}_{Sz+})$ and $\mathrm{ran}(\hat{P}_{Sx+})$ do not lie in a common invariant-subspace lattice given that $\mathrm{ran}(\hat{P}_{Sz+}) \in \mathcal{L}(\Sigma_{Sz})$ and $\mathrm{ran}(\hat{P}_{Sx+}) \in \mathcal{L}(\Sigma_{Sx})$. As a result, the proposition $P_{Sx+}$ represented by the subspace $\mathrm{ran}(\hat{P}_{Sx+})$ cannot assume any value in the state $|\Psi_{Sz+}\rangle \in \mathrm{ran}(\hat{P}_{Sz+})$, namely,\smallskip

\begin{equation}  
   b
   \Big(
      P_{Sx+}(|\Psi_{Sz+}\rangle)
   \Big)
   =
   b
   \left(
      \mathrm{Prop}
      \left(
         |\Psi_{Sz+}\rangle
         \in
         \mathrm{ran}(\hat{P}_{Sz+})
         \;\cancel{\;\wedge\;}\;
         \mathrm{ran}(\hat{P}_{Sx+})
      \right)
   \right)
   =
   \frac{0}{0}
   \;\;\;\;  .
\end{equation}
\smallskip

\noindent However, at some $k$ along the chain $\mathcal{H}_{S} \bigotimes_{k=1}^{N} \mathcal{H}_{k}$ (assuming that $N$ is large enough) the subspaces $\mathrm{ran}(\hat{P}_{Sz+})$ and $\mathrm{ran}(\hat{P}_{Sx+})$ will be the factors of the tensor products belonging to a common invariant-subspace lattice imposed on the subspaces of $\mathcal{H}_{SE}$. For the sake of simplicity, assume that this happens at $k = 1$, i.e.,\smallskip

\begin{equation}  
   \mathcal{H}_{SE}
   =
   \mathrm{ran}(\hat{1})
   =
   \mathcal{H}_{S}
   \otimes
   \mathrm{ran}(\hat{P}_{1z+})
   +
   \mathcal{H}_{S}
   \otimes
   \mathrm{ran}(\hat{P}_{1z-})
   \;\;\;\;  ,
\end{equation}
\smallskip

\noindent where\smallskip

\begin{equation}  
   \mathcal{H}_{S}
   \otimes
   \mathrm{ran}(\hat{P}_{1z+})
   =
   \left(
      \sum_{\alpha = \pm}
      \mathrm{ran}(\hat{P}_{Sx\alpha})
   \right)
   \otimes
   \mathrm{ran}(\hat{P}_{1z+})
   \;\;\;\;  ,
\end{equation}

\noindent \begin{equation}  
   \mathcal{H}_{S}
   \otimes
   \mathrm{ran}(\hat{P}_{1z-})
   =
   \left(
      \sum_{\alpha = \pm}
      \mathrm{ran}(\hat{P}_{Sz\alpha})
   \right)
   \otimes
   \mathrm{ran}(\hat{P}_{1z-})
   \;\;\;\;  .
\end{equation}
\smallskip

\noindent Now, the subspaces $\mathrm{ran}(\hat{P}_{Sz+})$ and $\mathrm{ran}(\hat{P}_{Sx+})$ are together in the invariant-subspace lattice $\mathcal{L}(\Sigma_{A})$ of the context $\Sigma_{A}$ associated with the composite system of two qubits:\smallskip

\begin{equation}  
   \mathrm{ran}(\hat{P}_{Sz+})
   \otimes
   \mathrm{ran}(\hat{P}_{1z-})
   \;
   ,
   \:
   \mathrm{ran}(\hat{P}_{Sx+})
   \otimes
   \mathrm{ran}(\hat{P}_{1z+})
   \:
   \in
   \mathcal{L}(\Sigma_{A})
   \;\;\;\;  ,
\end{equation}
\smallskip

\noindent where\smallskip

\begin{equation}  
   \Sigma_{A}
   =
   \left\{
      \hat{P}_{Sz+}
      \otimes
      \hat{P}_{1z-}
      \,
      ,
      \:
      \hat{P}_{Sz-}
      \otimes
      \hat{P}_{1z-}
      \,
      ,
      \:
      \hat{P}_{Sx+}
      \otimes
      \hat{P}_{1z+}
      \,
      ,
      \:
      \hat{P}_{Sx-}
      \otimes
      \hat{P}_{1z+}
   \right\}
   \;\;\;\;  .
\end{equation}
\smallskip

\noindent So, the proposition ``Spin of the qubit $S$ along the $x$-axis is $+\frac{\hbar}{2}$ \textit{and} spin of the environmental qubit along the $z$-axis is $+\frac{\hbar}{2}\:$'', denoted as $P_{Sx+}\!\wedge\! P_{1z+}$ and represented by the subspace $\mathrm{ran}(\hat{P}_{Sz+}) \otimes \mathrm{ran}(\hat{P}_{1z+})$, is \textit{determined} in the composite state describing the entanglement between two qubits $|\Psi_{Sz+}\rangle|\Psi_{1z-}\rangle \in \mathrm{ran}(\hat{P}_{Sz+}) \otimes \mathrm{ran}(\hat{P}_{1z-})$ (where $|a\rangle|b\rangle$ stands, as it is customary, for the tensor product $|a\rangle \otimes |b\rangle$). Concretely, since\smallskip

\begin{equation}  
   \mathrm{ran}(\hat{P}_{Sz+})
   \otimes
   \mathrm{ran}(\hat{P}_{1z-})
   \wedge
   \mathrm{ran}(\hat{P}_{Sx+})
   \otimes
   \mathrm{ran}(\hat{P}_{1z+})
   =
   \{0\}
   \;\;\;\;  ,
\end{equation}
\smallskip

\noindent the proposition $P_{Sx+}\!\wedge\! P_{1z+}$ assumes the value of falsity in the given state:

\begin{equation}  
   b
   \Big(
         P_{Sx+}\!\wedge\! P_{1z+}
         \:
         \big(|\Psi_{Sz+}\rangle|\Psi_{1z-}\rangle\big)
   \Big)
   =
   b
   \Big(
      \mathrm{Prop}
      \big(
         |\Psi_{Sz+}\rangle|\Psi_{1z-}\rangle
         \in
         \{0\}
      \big)
   \Big)
   =
   0
   \;\;\;\;  .
\end{equation}
\smallskip

\noindent On the other hand, the fact that the environmental qubits have the preferred set of states $\{|\Psi_{kz\pm}\rangle\}$ implies that the proposition $P_{1z+}$ represented by the subspace $\mathrm{ran}(\hat{P}_{1z+})$ has the value of falsity in the composite state $|\Psi_{Sz+}\rangle|\Psi_{1z-}\rangle$. As both $P_{Sx+}\!\wedge\! P_{1z+}$ and $P_{1z+}$ are false, the proposition $P_{Sx+}$ appears bivalent, i.e., either true or false, in this state:\smallskip

\begin{equation}  
   b
   \Big(
         P_{Sx+}
         \:
         \big(|\Psi_{Sz+}\rangle|\Psi_{1z-}\rangle\big)
   \Big)
   \in
   \{0,1\}
   \;\;\;\;  .
\end{equation}
\smallskip

\begin{figure}[ht!]
   \centering
   \includegraphics[scale=0.55]{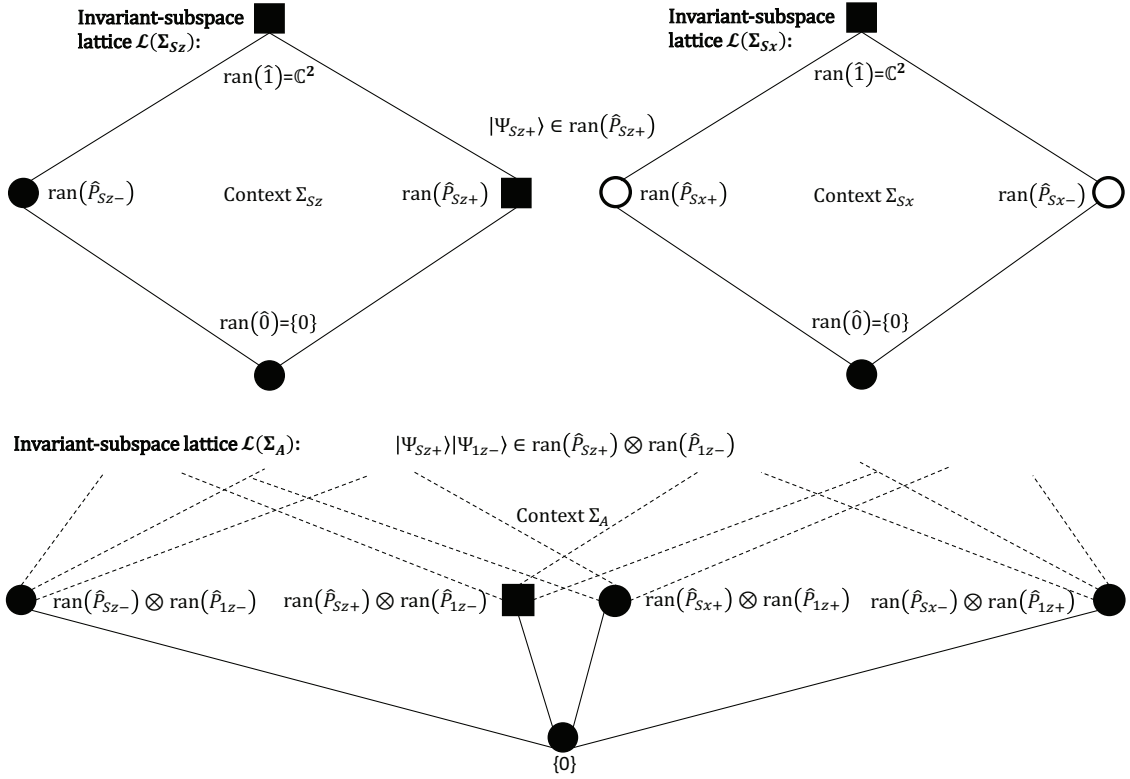}
   \caption{The bivaluation relation for the systems of one and two qubits\label{fig1}}
\end{figure}

\noindent The structures of the invariant-subspace lattices $\mathcal{L}(\Sigma_{Sz})$ and $\mathcal{L}(\Sigma_{Sx})$ as well as of $\mathcal{L}(\Sigma_{A})$, depicted in the form of \textit{modified Hasse diagrams}, are displayed in Figure \ref{fig1}.\\

\noindent Recall that a Hasse diagram is a type of mathematical diagram where each subspace corresponds to a vertex in the plane connected with another vertex by a line segment which goes upward from $\mathcal{H}^{\prime}$ to $\mathcal{H}^{\prime\prime}$ whenever $\mathcal{H}^{\prime} \subseteq \mathcal{H}^{\prime\prime}$ and there is no $\mathcal{H}^{\prime\prime\prime}$ such that $\mathcal{H}^{\prime} \subseteq \mathcal{H}^{\prime\prime\prime} \subseteq \mathcal{H}^{\prime\prime}$ \cite{Kiena}. Besides the information on the transitive reduction, the modified Hasse diagram shows the truth values of the propositions relating to the quantum system in the state $|\Psi\rangle$ by picturing the vertices that represent these propositions in the following way: the vertex is drawn as a black square if the proposition is true in $|\Psi\rangle$, the vertex is drawn as a black circle if the proposition is false in $|\Psi\rangle$, and the vertex is drawn as a hollow circle if the proposition cannot be described as either true or false in $|\Psi\rangle$.\\

\noindent In the upper half of the Figure \ref{fig1}, the truth values are given in the pure state $|\Psi_{Sz+}\rangle \in \mathrm{ran}(\hat{P}_{Sz+})$, while in the lower half of this Figure the truth values are given in the pure state $|\Psi_{Sz+}\rangle|\Psi_{1z-}\rangle\in \mathrm{ran}(\hat{P}_{Sz+}) \otimes \mathrm{ran}(\hat{P}_{1z-})$.\\

\noindent This model motivates the following.\\

\noindent Given that the interaction with the environment is, for all practical purposes, unavoidable, any quantum system $S$ is entangled with its environment $E$ such that the Hilbert space of the total system $SE$ is the tensor product\smallskip

\begin{equation}  
   \mathcal{H}_{SE}
   =
   \mathrm{ran}(\hat{1})
   =
   \mathcal{H}_{S}
   \otimes
   \mathcal{H}_{E}
   =
   \mathcal{H}_{S}
   \bigotimes_{k=1}^{N}
   \mathcal{H}_{k}
   \;\;\;\;  ,
\end{equation}
\smallskip

\noindent where $\mathcal{H}_{S}$ and $\mathcal{H}_{E}$ denote the Hilbert spaces of the system $S$ and environment $E$, while $\mathcal{H}_{k}$ stand for the Hilbert spaces of the systems $E_k$ composing the environment $E$.\\

\noindent Suppose that the nontrivial subspaces $\mathrm{ran}(\hat{P})$ and $\mathrm{ran}(\hat{Q})$ of system's Hilbert space $\mathcal{H}_{S}$ do not belong to one context when treated in isolation, that is,\smallskip

\begin{equation}  
   \mathcal{H}_{S}
   =
   \mathrm{ran}(\hat{1})
   =
   \mathrm{ran}(\hat{P})
   +
   \mathrm{ran}(\hat{P}^{\,\prime})
   +
   \dots
   =
   \mathrm{ran}(\hat{Q})
   +
   \mathrm{ran}(\hat{Q}^{\,\prime})
   +
   \dots
   =
   \dots
   \;\;\;\;  ;
\end{equation}
\smallskip

\noindent thus, the proposition $Q$ represented by the subspace $\mathrm{ran}(\hat{Q})$ has the truth-value gap in the state $|\Psi\rangle \in \mathrm{ran}(\hat{P})$, meaning that $b\big(Q(|\Psi\rangle)\big) =\frac{0}{0}$.\\

\noindent After the interaction between the system and the environment, these subspaces turn into factors of the tensor products $\mathrm{ran}(\hat{P}) \otimes \mathcal{H}_E^{\prime}$ and $\mathrm{ran}(\hat{Q}) \otimes \mathcal{H}_E^{\prime\prime}$, where $\mathcal{H}_E^{\prime}$ and $\mathcal{H}_E^{\prime\prime}$ are the nontrivial subspaces of $\mathcal{H}_E$. Because of an enormous number of degrees of freedom in the environment, the chain $\mathcal{H}_{S} \bigotimes_{k=1}^{N} \mathcal{H}_{k}$ is \textit{expected to be large enough} to put $\mathrm{ran}(\hat{P}) \otimes \mathcal{H}_E^{\prime}$ and $\mathrm{ran}(\hat{Q}) \otimes \mathcal{H}_E^{\prime\prime}$ together in the invariant-subspace lattice $\mathcal{L}(\Sigma_{SE})$ of the context $\Sigma_{SE}$ associated with the total system $SE$, namely,\smallskip

\begin{equation}  
   \Sigma_{SE}
   =
   \left\{
      \dots
      \,
      ,
      \:
      \hat{P}
      \otimes
      \hat{E}^{\,\prime}
      \,
      ,
      \:
      \dots
      \,
      ,
      \:
      \hat{Q}
      \otimes
      \hat{E}^{\,\prime\prime}
      \,
      ,
      \:
      \dots
   \right\}
   \;\;\;\;  ,
\end{equation}
\smallskip

\noindent where $\hat{E}^{\,\prime}$ and $\hat{E}^{\,\prime\prime}$ denote the projection operators corresponding exactly to the subspaces $\mathcal{H}_E^{\prime}$ and $\mathcal{H}_E^{\prime\prime}$. One can infer from this that after the interaction between $S$ and $E$, the proposition $Q$ becomes bivalent.\\

\noindent To be sure, let the subspaces $\mathcal{H}_E^{\prime}$ and $\mathcal{H}_E^{\prime\prime}$ represent correspondingly the propositions $E^{\,\prime}$ and $E^{\,\prime\prime}$, which have the value of truth in the environmental states $|\epsilon^{\,\prime}\rangle \in \mathcal{H}_E^{\prime}$ and $|\epsilon^{\,\prime\prime}\rangle \in \mathcal{H}_E^{\prime\prime}$ in that order. According to \textit{the stability criterion} \cite{Schlosshauer}, the environment $E$ has the preferred set of states, in which the correlation between any two environmental systems $E_k$ and $E_{m \neq k}$ is left undisturbed by the subsequent formation of correlations with other systems of the environment $E$. This implies that in the pure state $|\Psi\rangle|\epsilon^{\,\prime}\rangle \in \mathrm{ran}(\hat{P}) \otimes \mathcal{H}_E^{\prime}$ describing the entanglement between the system $S$ and its environment $E$, the proposition $E^{\,\prime\prime}$ has the value of falsity, i.e., $b\big(E^{\,\prime\prime}(|\Psi\rangle|\epsilon^{\,\prime}\rangle)\big) = 0$. Then, seeing as the proposition $Q \wedge E^{\,\prime\prime}$ represented by subspace $\mathrm{ran}(\hat{Q}) \otimes \mathcal{H}_E^{\prime\prime}$ also assumes the value of falsity in this state, explicitly,\smallskip

\begin{equation}  
   b
   \Big(
      Q \wedge E^{\,\prime\prime}
      \left(
         |\Psi\rangle|\epsilon^{\,\prime}\rangle
      \right)
   \Big)
   =
   b
   \bigg(\!
      \mathrm{Prop}
      \Big(
         |\Psi\rangle|\epsilon^{\,\prime}\rangle
         \in
         \smash{\underbrace{
            \mathrm{ran}(\hat{P}) \otimes \mathcal{H}_E^{\prime}
            \wedge
            \mathrm{ran}(\hat{Q}) \otimes \mathcal{H}_E^{\prime\prime}
         }_{
            =\{0\}
         }}
         \rule[-5.25ex]{0ex}{0ex}
      \,\Big)\!
   \bigg)
   =
   0
   \;\;\;\;  ,
\end{equation}
\smallskip

\noindent one can conclude that $b\big(Q(|\Psi\rangle|\epsilon^{\,\prime}\rangle)\big) \in \{0,1\}$.\\

\noindent Hence, the environment induces bivalence; in other words, the interaction between the quantum system and the environment brings on the transition from supervaluation semantics to a bivalent semantics:\smallskip

\begin{equation}  
   b
   \Big(
      Q(|\Psi\rangle)
   \Big)
   =
   \frac{0}{0}
   \;
   \to
   \;
   b
   \Big(
      Q(|\Psi\rangle|\epsilon^{\,\prime}\rangle)
   \Big)
   \in
   \{0,1\}
   \;\;\;\;  .
\end{equation}
\smallskip

\noindent In this view, the quantum-mechanical measurement (provided that any measuring device or apparatus acts as an environment) changes the state where $Q$ is neither true nor false into the state in which $Q$ is either true or false.\\

\section{The classical limit of quantum logic}  

\noindent The environmentally induced bivalence is worth comparing with the classical limit of quantum logic.\\

\noindent Let $\mathcal{L}(\mathbb{C}^2)$ denote the Hilbert lattice imposed on all closed linear subspaces of the Hilbert space $\mathcal{H} = \mathbb{C}^2$. Consider the sublattice of $\mathcal{L}(\mathbb{C}^2)$ whose nontrivial subspaces $\mathrm{ran}(\hat{P}_{q\pm})$ correspond to the projection operators $\hat{P}_{q±}$ for spin of the qubit along the $z$, $x$ and $y$ axes:\smallskip

\begin{equation}  
   \mathcal{K}(\mathbb{C}^2)
   =
   \left\{
      \{0\}
      \,
      ,
      \,
      \mathrm{ran}(\hat{P}_{z+})
      \,
      ,
      \,
      \mathrm{ran}(\hat{P}_{z-})
      \,
      ,
      \,
      \mathrm{ran}(\hat{P}_{x+})
      \,
      ,
      \,
      \mathrm{ran}(\hat{P}_{x-})
      \,
      ,
      \,
      \mathrm{ran}(\hat{P}_{y+})
      \,
      ,
      \,
      \mathrm{ran}(\hat{P}_{y-})
      \,
      ,
      \,
      \mathbb{C}^2
   \right\}
   \;\;\;\;  .
\end{equation}
\smallskip

\noindent As it is easily seen, the invariant-subspace lattices $\mathcal{L}(\Sigma_q) = \{ \{0\},\,\mathrm{ran}(\hat{P}_{q+}),\,\mathrm{ran}(\hat{P}_{q-}),\,\mathbb{C}^2\}$ are pasted together in $\mathcal{K}(\mathbb{C}^2)$ at the trivial subspaces $\{0\}$ and $\mathbb{C}^2$. Consequently, the lattices $\mathcal{L}(\Sigma_q)$ can be identified with Boolean blocks of the sublattice $\mathcal{K}(\mathbb{C}^2)$ (that can be called \textit{Hilbert sublattice}). In this sense, the Hilbert sublattice $\mathcal{K}(\mathbb{C}^2)$  is \textit{the pasting of a continuity of the Boolean blocks} $\mathcal{L}(\Sigma_q)$.\\

\noindent As a result, in $\mathcal{K}(\mathbb{C}^2)$ the meet operation exists for pairs of the nontrivial subspaces belonging to non-commutable projection operators (such as $\mathrm{ran}(\hat{P}_{z+})$ and $\mathrm{ran}(\hat{P}_{x+})$), and so the propositional function $\mathrm{Prop}$ is determined for such pairs. For this reason, within the structure of a Hilbert sublattice (and hence quantum logic), quantum propositions are always bivalent.\\

\noindent Recall that closed linear subspaces of a Hilbert space \textit{commute with each other} if the condition\smallskip

\begin{equation}  
   \mathcal{H}^{\prime}
   \,
   ,
   \mathcal{H}^{\prime\prime}
   \,
   \subseteq
   \mathcal{H}
   \;
   \implies
   \;
   \mathcal{H}^{\prime}
   \cap
   \left(
      \mathcal{H}^{\prime}
      \cap
      (\mathcal{H}^{\prime\prime})^{\perp}
   \right)^{\perp}
   \subseteq
   \mathcal{H}^{\prime\prime}
   \;\;\;\;   
\end{equation}
\smallskip

\noindent is applicable to them \cite{Pavicic}. It is readily to see that any pair of the subspaces in the Boolean blocks, i.e., $\mathcal{H}^{\prime}\, , \, \mathcal{H}^{\prime\prime} \in \mathcal{L}(\Sigma_q)$, adheres to this condition, but it is not applicable to a pair $\mathcal{H}^{\prime}\, , \, \mathcal{H}^{\prime\prime} \in \mathcal{L}(\mathcal{H})$ where $\mathcal{H}^{\prime} = \mathrm{ran}(\hat{P})$, $\mathcal{H}^{\prime\prime} = \mathrm{ran}(\hat{Q})$ and $\hat{P}\hat{Q} \neq \hat{Q}\hat{P}$.\\

\noindent Let the lack of commutativity of two projection operators $\hat{P}$ and $\hat{Q}$ be measured by their commutator, $\hat{C}$, defined by $\hat{C} = [\hat{P},\hat{Q}] = \hat{P}\hat{Q} - \hat{Q}\hat{P}$. Then, one can state that the transition from quantum logic to classical (Boolean) logic is the transition\smallskip

\begin{equation} \label{TRAN} 
   \hat{C}
   \neq
   0
   \;
   \to
   \;
   \hat{C}
   =
   0
   \;\;\;\;   
\end{equation}
\smallskip

\noindent for any $\hat{P}$ and $\hat{Q}$.\\

\noindent The difficulty with this transition is that it cannot be described by means of the Schr\"{o}dinger equation which determines the unitary (deterministic) evolution of a quantum system \cite{Losada}. And so, the transition (\ref{TRAN}) can only be explained through either a non-unitary time evolution or deformation quantization and the associated continuity between quantum and classical algebras of observables in the limit $\hbar \to 0$. The problem is that both mentioned options are rather questionable.\\

\noindent Indeed, a non-unitary time evolution implies a modification of the Schr\"{o}dinger equation (e.g., adding a nonlinear term into the equation \cite{Weber, Ghose}, allowing a non-Hermitian Hamiltonian in it \cite{Moiseyev}) equivalent to the recognition of the existence of fundamental processes not governed by standard quantum mechanics.\\

\noindent As to the second option, assume for simplicity that two incommensurable observables $p$ and $q$ have a discrete spectrum, and therefore their operators $\hat{p}$ and $\hat{q}$ are given by $\hat{p} = \sum_n p_n \hat{P}_n$ and $\hat{q} = \sum_m q_m \hat{Q}_m$, where $\hat{P}_n$ and $\hat{Q}_m$ denote projection operators, while $p_n$ and $q_m$ stand for eigenvalues of the operators $\hat{p}$ and $\hat{q}$. Consider the commutator $\hat{c}$ of $\hat{p}$ and $\hat{q}$:\smallskip

\begin{equation}  
   \hat{c}
   =
   [\hat{p},\hat{q}]
   =
   \sum_n
   \sum_m
   p_n
   q_m
   \hat{C}_{nm}
   \neq
   0
   \;\;\;\;  ,
\end{equation}
\smallskip

\noindent where $\hat{C}_{nm} = [\hat{P}_n,\hat{Q}_m]$. In accordance with the axioms of deformation quantization, the commutator $\hat{c}$ may be presented formally as\smallskip

\begin{equation}  
   \hat{c}
   =
   i\hbar
   \widehat{\{p,q\}}
   +
   O(\hbar^2)
   \;\;\;\;  ,
\end{equation}
\smallskip

\noindent where $\{p,q\}$ is equal to the Poisson bracket of the observables $p$ and $q$ \cite{Rosaler}. One can infer from this that the commutator $\hat{c}$ -- hence any commutator $\hat{C}_{nm}$ -- becomes zero in the limit $\hbar \to 0$. However, it is not clear if \textit{the deformation quantization problem} (i.e. the deformation of a Poisson algebra, which preserves its commutative subalgebra) \textit{has a generic solution} \cite{Sharygin}. Accordingly, it might be that only specific choices of the operators $\hat{p}$ and $\hat{q}$ are suitable for $\lim_{\,\hbar \to 0}\, \hat{c} = 0$.\\

\noindent To conclude, one can observe that in comparison with the environmentally induced bivalence, the classical limit of quantum logic seems more problematic.\\

\bibliographystyle{References}

\end{document}